\documentclass[aps,pra,superscriptaddress,showpacs,showkeys]{revtex4}
\usepackage{epsfig}
\usepackage{graphicx}
\usepackage{bm}
\usepackage{amssymb}
\usepackage{amsmath}
\newcommand{\txt}[1]{{\rm #1}}

\newcommand{\zh}{\bm}
\newcommand{\real}{\mathop{\rm Re}\nolimits}
\newcommand{\image}{\mathop{\rm Im}\nolimits}

\newcommand{\trace}{\mathop{\rm Tr}\nolimits}

\newcommand{\zhr}{{\zh r}}

\newcommand{\zhp}{{\zh p}}

\newcommand{\zhalpha}{{\zh\alpha}}

\newcommand{\zhnu}{{\zh\nu}}

\newcommand{\zhsigma}{{\zh\sigma}}
\newcommand{\zhzeta}{{\zh\zeta}}

\newcommand{\unin}{\hat{\zh n}}
\newcommand{\unik}{\hat{\zh k}}
\newcommand{\uniq}{\hat{\zh q}}

\newcommand{\zhgamma}{{\zh \gamma}}

\newcommand{\Eq}[1]{Eq.\ (\ref{#1})}

\newcommand{\Eqss}[2]{Eqs.\ (\ref{#1}),(\ref{#2})}
\newcommand{\Eqsss}[3]{Eqs.\ (\ref{#1}),(\ref{#2}),(\ref{#3})}
\newcommand{\Eqsamb}[2]{Eqs.\ (\ref{#1}-\ref{#2})}
\newcommand{\Fig}[1]{Fig.\ \ref{#1}}

\newcommand{\matrixb}[4]{
	\left(
	\begin{array}{cc}
		#1&#2\\
		#3&#4\\
	\end{array}
	\right)
}

\begin{document}
    	\title{Resonant elastic scattering of polarized electrons on H-like ions}
    	\author{ D.\ M.\ Vasileva}
    	\affiliation{Department of Physics,
    		St.\ Petersburg State University, 7/9 Universitetskaya nab.,
    		St. Petersburg, 199034, Russia}
    	\author{ K.\ N.\ Lyashchenko}
    	\affiliation{Institute of Modern Physics, Chinese Academy of Sciences, Lanzhou 730000, China}
    	\author{ A. B. Voitkiv }
    	\affiliation{Institute for Theoretical Physics I,
    		Heinrich-Heine-University of D\"usseldorf, D\"usseldorf 40225, Germany}
    	\author{D. Yu}
    	\affiliation{Institute of Modern Physics, Chinese Academy of Sciences, Lanzhou 730000, China}
    	\affiliation{University of Chinese Academy of Sciences, Beijing 100049, China}
    	\author{O.\ Yu.\ Andreev}
    	\affiliation{Department of Physics,
    		St.\ Petersburg State University, 7/9 Universitetskaya nab.,
    		St. Petersburg, 199034, Russia}
    	\affiliation{Petersburg Nuclear Physics Institute named by B.P. Konstantinov of National
    		Research Centre "Kurchatov Institute", Gatchina, Leningrad District 188300, Russia}
    	\date{\today}
    		
    \begin{abstract}
    	The polarization properties of the elastic electron scattering on H-like ions are investigated within the framework of the relativistic QED theory. The polarization properties are determined by a combination of relativistic effects and spin exchange between the incident and bound electrons. The scattering of a polarized electron on an initially unpolarized ion is fully described by five parameters. We study these parameters for non-resonant scattering, as well as in the vicinity of LL resonances, where scattering occurs through the formation and subsequent decay of intermediate autoionizing states. The study was carried out for ions from $\txt{B}^{4+}$ to $\txt{Xe}^{53+}$. Special attention was paid to the study of asymmetry in electron scattering.
    \end{abstract}
    \maketitle{}
\section{Introduction}
The scattering of an unpolarized electron by the Coulomb field of an atomic nucleus has axial symmetry with respect to the direction of the momentum of the incident electron.
In 1929, N. F. Mott suggested that in the scattering by the Coulomb field an initially unpolarized electron becomes partially polarized and the resulting polarization is directed perpendicular to the plane of scattering (and, therefore, changes depending on the direction of the scattered electron momentum) \cite{mott29}. It was also noted that if the electron polarized in this way is then scattered on the Coulomb potential again, the differential cross section of this process is no longer symmetric in respect to the direction of the incident electron momentum and, therefore, the polarization of the electron can be detected experimentally in such scattering. The cause of these effects is most easily seen in the electron rest frame. When the electron is moving in the electric Coulomb field, in the rest frame of the electron there is magnetic field that interacts with the magnetic moment of the electron. This interaction explains the appearance of asymmetry in the electron scattering.

The Mott prediction was confirmed in double scattering experiment in 1943 \cite{Shull1943}. Since then, the asymmetry arising in Mott scattering has been utilized to produce and measure electron polarization. Today Mott polarimeters are widely used in different fields of physics such as atomic, nuclear or particle physics \cite{Gay1992,Roca_Maza2017}.

A general theory for describing the electron polarization is presented in \cite{Tolhoek1954, kessler}. A detailed study of the polarization properties of the Coulomb scattering is given in \cite{Tolhoek1956,johnson1961}. The asymmetry function that describes both the polarization acquired by the electron during scattering and the possible asymmetry of the differential cross section was of most interest. This function also known as Sherman asymmetry function was thoroughly investigated for the Coulomb scattering in \cite{johnson1961,sherman56}.

Sherman asymmetry function can also be introduced for the electron scattering on atoms or molecules. The polarization properties of the electron scattering were studied for collisions with a wide range of neutral atoms \cite{Lin1963,Cox1967,Gellrich91,Dummler_1995,Dorn1998,Ahmed2007,Haque2017,Jakubassa2018107}. Several codes for calculation of polarization parameters were suggested \cite{Salvat2005,DRAGOWSKI201648,Dapor2018,KOKKORIS201944}.
	
A general theory for the electron scattering on H-like ions was developed in \cite{Burke1974}. In this case, the polarization of the initial electron can change not only due to the Coulomb scattering on the Coulomb potential of the ion, but also due to the spin exchange with a bound electron.

The electron scattering on ions was studied experimentally in \cite{huber1994PhysRevLett.73.2301,greenwood1995PhysRevLett.75.1062,troth1996PhysRevA.54.R4613,zouros2003PhysRevA.68.010701,benis2004pra69-052718,benis2006PhysRevA.73.029901}. The focus of these experiments was on obtaining the differential cross sections.

Resonant scattering of electrons on H-like ions is the simplest process in which Coulomb scattering, Auger decay and their interference can be equally important. If the energy of the initial state ($\varepsilon(e)$ + $\varepsilon(1s)$) is close to the energy of
one of the autoionizing states, then the scattering of an electron can occur due to the formation of a doubly excited state (d) and its subsequent Auger decay:
\begin{equation}
e^-+X^{(Z-1)+}(1s)
\to
X^{(Z-2)+}(d)
\,\to\,
e^-+
X^{(Z-1)+}(1s)
\,. 
\end{equation}
In the present paper we consider the impact energies for which LL autoionizing states (for instance, $(2s2s)$) are of importance.

The formation of autoionizing states is manifested in the appearance of a clear resonance structure of the cross section as a function of the incident electron energy.
The resonances in the cross section mainly have a Fano-like shape due to the strong interference between the resonance channel and the Coulomb scattering channel.
In our previous paper \cite{res2020} we developed a QED theory for the description of such a process. 

In the present work, we apply the methods of describing the polarization properties developed in \cite{Tolhoek1954} and \cite{Burke1974} to the process of elastic scattering of electrons on H-like ions. We introduce and investigate five parameters that determine the polarization properties of the scattering of electrons on unpolarized ions. Special attention is paid to the study of the asymmetry function of both the scattered electron and the ion. The study was carried out for the resonant energy region of incident electrons.

\section{Scattering of polarized electrons}
A partially polarized electron beam is described by the density matrix $\hat{\rho}_s$ \cite{akhiezer65b}:
\begin{eqnarray}
\hat{\rho}_s=\dfrac{1}{2}(m+\hat{p})(1-\gamma_5\hat{s})\,, \label{fulldm}
\end{eqnarray}
where
\begin{eqnarray}
&\hat{p}=p_\mu\gamma^\mu=p_0\gamma_0-\zhp\zhgamma, &\\
&\hat{s}=s_\mu\gamma^\mu=s_0\gamma_0-\zh{s}\zhgamma, &\\
&\zh{s}=\zhzeta+\dfrac{(\zhzeta\zhp)\zhp}{m(\varepsilon+m)},s_0=\dfrac{(\zhzeta\zhp)}{m}\,,& \label{fulldmend}
\end{eqnarray}
$m$ is the electron mass, $\zhzeta$ is the polarization vector of the electron in the electron rest frame, $P=|\zhzeta|\leqslant1$ - degree of polarization. Dirac gamma matrices here are defined as:
\begin{eqnarray}
\gamma_0&=&\beta=\matrixb{I}{0}{0}{-I}, \\
\zhgamma&=&\beta\zhalpha=\matrixb{0}{\zhsigma}{-\zhsigma}{0},\\
\gamma_5&=&i\gamma_1\gamma_2\gamma_3\gamma_0\,. 
\end{eqnarray}
 
If the density matrix of the initial state is $\hat{\rho_i}$, then the density matrix describing the electron after scattering can be written as:
\begin{equation}
\hat{\rho}=S\hat{\rho_i}\bar{S},
\end{equation}
where $S$ is the $S$-matrix, $\bar{S}=\gamma_0 S^{+} \gamma_0$.

If one is interested in the probability of transition from state $i$ described by the density matrix $\hat{\rho_i}$ into state $f$ described by $\hat{\rho_f}$, one should use the following formula for the differential cross section with the appropriate coefficient \cite{Tolhoek1954}:
\begin{equation}
d\sigma(\zhp_i,\zhzeta_i;\zhp_f,\zhzeta_f)\sim\trace(\hat{\rho_f}\hat{\rho})=\trace(\hat{\rho_f}S\hat{\rho_i}\bar{S})\,.  
\end{equation}
We note that here $\hat{\rho_f}$ is the density matrix that determines the state of the electron detected in the experiment rather than the state of the scattered electron.

In the case of the elastic electron scattering on an hydrogen-like ion  we must handle the two-electron system. Since before the scattering we can consider incident and bound electrons as non-interacting, the density matrix of the initial state is simply a direct product of corresponding single-electron density matrices. Furthermore, the polarization of incident, scattered and $1s$ electrons can be fully described by spinors alone. Therefore, it is sufficient to use the $4\times4$ density matrices of form:
\begin{equation}
\hat{\rho}_{\zhzeta \zh{\eta}}=\dfrac{1}{4}(1+\zhzeta\zhsigma_1)(1+\zh{\eta}\zhsigma_2)\,,
\end{equation} 
where $\zhzeta$ is the polarization vector of incident (or scattered electron), $\zh{\eta}$ is the polarization vector of the ion in $1s$ state, matrix indices $1$ and $2$ refer to the incident (scattered) and $1s$ electrons respectively.

It is convenient to take $z$-axis in the direction of the incident beam and introduce angles $\theta$ and $\varphi$ for the direction of the scattered electron momentum.

The initial and final states of the two-electron system are described by:
\begin{eqnarray}
\Psi^i_{m_i,\mu_i}(\zhp_i)&=&\dfrac{1}{\sqrt{2}}\det{\lbrace\psi^{(+)}_{\zhp_i\mu_i}(\zhr_1)\psi_{1s_{m_i}}(\zhr_2)\rbrace} \label{psii}\,,\\
\Psi^f_{m_f,\mu_f}(\zhp_f)&=&\dfrac{1}{\sqrt{2}}\det{\lbrace\psi^{(-)}_{\zhp_f\mu_f}(\zhr_1)\psi_{1s_{m_f}}(\zhr_2)\rbrace} \label{psif}\,,
\end{eqnarray}
where $m_i$ and $m_f$ are $1s$ electron spin projections on $z$-axis and $\psi^{(\pm)}_{\zhp\mu}(\zhr)$ is in- $(+)$ or out-going $(-)$ wave function of an electron in an external Coulomb field with an asymptotic
momentum ($\zhp=p \hat{\zhnu}$) and polarization ($\mu$):
\begin{eqnarray}
\psi^{(\pm)}_{\zhp\mu}(\zhr)=\dfrac{(2\pi)^{3/2}}{\sqrt{p\varepsilon}}\sum_{jlm}\Omega^{+}_{jlm}(\zhnu)v_{\mu}(\zhnu)e^{\pm i\phi_{jl}} i^l \psi_{\varepsilon jlm}(\zhr)\,.
\end{eqnarray}
Spinor $v_{\mu_{\zhzeta}}$ describes the electron with the (asymptotic) projection of spin $\mu_{\zhzeta}$ on the direction $\hat{\zhzeta}=\zhzeta/|\zhzeta|$ and is determined by the following equation:
\begin{equation}
\dfrac{1}{2}(\hat{\zhzeta}\zhsigma)v_{\mu_{\zhzeta}}=\mu_{\zhzeta}v_{\mu_{\zhzeta}}\,. 
\end{equation}

The scattering is described by $4\times4$ matrix $M$: 
\begin{eqnarray}
M_{m_i,\mu_i;m_f,\mu_f}(\theta,\varphi)&=&U_{m_i,\mu_i;m_f,\mu_f}(\theta,\varphi) \label{mcalcbegin}\\
&=&U^{\txt{Coul}}_{\mu_i\mu_f}(\theta,\varphi)\delta_{m_i m_f}+U^{\txt{Auger}}_{m_i,\mu_i;m_f,\mu_f}(\theta,\varphi)\,.
\end{eqnarray}

It is convenient to consider the total amplitude of the process as the sum of Coulomb amplitude corresponding to the Coulomb scattering on the screened potential of the nucleus $(Z-1)/r$ and Auger amplitude. Auger amplitude includes contributions from the interelectron interaction: the resonant channel as well as the remaining part of the non-resonant channel. 

The Coulomb term for the scattering matrix $M$ has the form \cite{burke2011b}:
\begin{eqnarray}
&M^{\txt{Coul}}_{m_i,\mu_i;m_f,\mu_f}(\theta,\varphi)=\delta_{m_i m_f}\mathcal{M}^{\txt{Coul}}_{\mu_i\mu_f}(\theta,\varphi)\,,& \label{Couldef1}\\
&\mathcal{M}^{\txt{Coul}}_{\mu_i\mu_f}(\theta,\varphi)=\matrixb{f(\theta)}{g(\theta)e^{-i\varphi}}{-g(\theta)e^{i\varphi}}{f(\theta)}\, .&  \label{Couldef2} 
\end{eqnarray}   
The resonance term $U^{\txt{Auger}}_{m_i,\mu_i;m_f,\mu_f}$ is given by \cite{res2020}:
\begin{equation}
U^{\txt{Auger}}_{m_i,\mu_i;m_f,\mu_f}=\langle\Psi^{f}_{m_f,\mu_f}(\zhp_f)\vert\Delta\hat{V}\vert\Phi_{m_i,\mu_i}(\zhp_i)\rangle\,, \label{mcalcend}
\end{equation}  
where $\Delta\hat{V}$ is derived within the QED perturbation theory order by order, $\Phi_{m_i,\mu_i}(\zhp_i)$ is constructed from $\Psi^i_{m_i,\mu_i}(\zhp_i)$ within the line-profile approach \cite{andreev08pr}, $\Psi^i_{m_i,\mu_i}(\zhp_i)$ and $\Psi^{f}_{m_f,\mu_f}(\zhp_f)$ are given by \Eqss{psii}{psif} with polarization projections $\mu_i$ and $\mu_f$ taken with respect to the $z$-axis.

Scattering matrix $M$ can be expressed in terms of Pauli matrices $\hat{\sigma}$. Since $M$ must be invariant in respect to space rotation and reflection and also time reversal, only six linearly independent terms remain \cite{Burke1974}:
\begin{eqnarray}
M=a_0+a_1(\zhsigma_1\unin)+a_2(\zhsigma_2\unin)+a_{12}(\zhsigma_1\unin)(\zhsigma_2\unin)\nonumber\\
+b_{12}(\zhsigma_1\unik)(\zhsigma_2\unik)+c_{12}(\zhsigma_1\uniq)(\zhsigma_2\uniq)\,. \label{Mexpansion}
\end{eqnarray}
Here vectors $\unin$, $\unik$, $\uniq$ are introduced:
\begin{eqnarray}
\unin&=&\dfrac{\zhp_i\times\zhp_f}{|\zhp_i\times\zhp_f|}\,,\\
\unik&=&\dfrac{\zhp_i+\zhp_f}{|\zhp_i+\zhp_f|}\,,\\
\uniq&=&\dfrac{\zhp_i-\zhp_f}{|\zhp_i-\zhp_f|}\,,
\end{eqnarray}
and $a_0$, $a_1$, $a_2$, $a_{12}$, $b_{12}$, $c_{12}$ are functions of only $\theta$.

For the Coulomb scattering:
\begin{equation}
M^{\txt{Coul}}=f(\theta)+ig(\theta)(\zhsigma_1\unin)\,.
\end{equation}

We note that the scattering can be fully described with eleven real numbers since matrix $M$ is determined by 6 complex numbers (\Eq{Mexpansion}) and the phase of $M$ can be chosen arbitrarily \cite{Burke1974}. 

The density matrix of the two-electron system after scattering with the initial state described by $\hat{\rho_i}$ is given by:
\begin{equation}
\hat{\rho}=M\hat{\rho_i}M^+\,.
\end{equation}

If after scattering electron and ion in the state described by $\hat{\rho_f}$ are detected, the corresponding differential cross section can be obtained with the use of the following formula:
\begin{equation}
d\sigma(\zhp_i,\zhzeta_i,\zh{\eta}_i;\zhp_f,\zhzeta_f,\zh{\eta}_f)=\dfrac{2\pi}{j}\trace(\hat{\rho_f}M\hat{\rho_i}M^{+}) \delta(\varepsilon_f-\varepsilon_i) \dfrac{d^3\zhp_f}{(2\pi)^3}\,, \label{sigma}
\end{equation}
\begin{equation}
j=\dfrac{p_i}{\varepsilon_i}\,,
\end{equation}
\begin{eqnarray}
\hat{\rho}_i&=&\dfrac{1}{4}(1+\zhzeta_i\zhsigma_1)(1+\zh{\eta}_i\zhsigma_2)\,,\label{mi}\\
\hat{\rho}_f&=&\dfrac{1}{4}(1+\zhzeta_f\zhsigma_1)(1+\zh{\eta}_f\zhsigma_2)\,.\label{mf}
\end{eqnarray}

If an electron is scattered on an unpolarized ion and only electron polarization is of interest, in order to obtain the respective differential cross section one should set $\zh{\eta}_i$ and $\zh{\eta}_f$ in \Eqsamb{sigma}{mf} to zero. After calculating the trace in \Eq{sigma} we get the following formula:
\begin{eqnarray}
\dfrac{d\sigma}{d\Omega}(\zhp_i,\zhzeta_i;\zhp_f,\zhzeta_f)&=&\dfrac{1}{4}w\lbrace I(\theta)(1+\zhzeta_i\zhzeta_f)+D(\theta)(\zhzeta_i\unin+\zhzeta_f\unin) \nonumber \\
&&+F(\theta)[\zhzeta_i\times\zhzeta_f]\unin-G(\theta)[\zhzeta_i\times\unin][\zhzeta_f\times\unin] \nonumber\\
&&-H(\theta)[\zhzeta_i\times\unik][\zhzeta_f\times\unik]-K(\theta)[\zhzeta_i\times\uniq][\zhzeta_f\times\uniq]\rbrace\,, \label{cs}
\end{eqnarray} 
where
\begin{eqnarray}
I(\theta)&=&|a_0|^2+|a_1|^2+|a_2|^2+|a_{12}|^2+|b_{12}|^2+|c_{12}|^2\,,\\
D(\theta)&=&2\real(a_0 a_1^*+a_2 a_{12}^*)\,,\\ \label{f1}
F(\theta)&=&2\image(a_0 a_1^*+a_2 a_{12}^*)\,,\\
G(\theta)&=&2(|a_1|^2+|a_{12}|^2)\label{fg}\,,\\
H(\theta)&=&2|b_{12}|^2 \label{fh}\,,\\ 
K(\theta)&=&2|c_{12}|^2 \label{f2}\,,\\ 
w&=&\dfrac{2\pi}{j} \int d\varepsilon_f \delta(\varepsilon_f-\varepsilon_i) \dfrac{\varepsilon_f p_f}{(2\pi)^3}=\dfrac{\varepsilon_i^2}{(2\pi)^2}\,.
\end{eqnarray} 
Similar formula for the Coulomb scattering was derived in \cite{johnson1961,Tolhoek1956}.

If an unpolarized electron is scattered on an unpolarized ion, density matrix of the two-electron system after scattering is:
\begin{equation}
\hat{\rho}=\dfrac{1}{4}MM^+=\tilde{I}(\theta)\left(1+\zhzeta^e(\theta)\zhsigma_1+\zh{\eta}^{ion}(\theta)\zhsigma_2+\sum_{ij}C_{ij}(\theta)\sigma_{1i}\sigma_{2j}\right)\,,
\end{equation}
where $\zhzeta^e$ - polarization vector of the scattered electron, $\zh{\eta}^{ion}$ - polarization vector of the ion after scattering, $C_{ij}$ determines the correlation between polarizations of the electron and the ion.
\begin{eqnarray}
\tilde{I}(\theta)&=&\dfrac{1}{4}\trace(MM^+)=I(\theta)\,,\\
\zhzeta^{e}&=&\dfrac{\trace(MM^+\zhsigma_1)}{\trace(MM^+)}=\unin\dfrac{D(\theta)}{I(\theta)} \label{zetaD} \label{finpol}\,,\\
\zh{\eta}^{\txt{ion}}&=&\dfrac{\trace(MM^+\zhsigma_2)}{\trace(MM^+)}=\unin\dfrac{D^{\txt{ion}}(\theta)}{I(\theta)}\,,\\
D^{\txt{ion}}(\theta)&=&2\real(a_0 a_2^*+a_1 a_{12}^*)\,.
\end{eqnarray}

Functions $D$ and $D^{\txt{ion}}$ determine the transverse polarization of the electron and the ion acquired in the process of scattering.
The polarization vector of the scattered electron can also be presented as the difference between cross sections for scattering of an unpolarized electron on an unpolarized ion with opposite scattered electron spin projections on $\unin$ ($\mu_{\perp}=\pm 1/2$) divided by the total cross section: 
\begin{eqnarray}
\zhzeta^{e}&=&\unin \cdot S_{0\unin} \label{zetaS}\,,
\end{eqnarray}
where
\begin{eqnarray}
&&S_{\hat{\zhzeta_i}\hat{\zhzeta_f}}=\dfrac{\left( \dfrac{d\sigma}{d\Omega}\right) ^+_{\hat{\zhzeta_i}\hat{\zhzeta_f}}-\left( \dfrac{d\sigma}{d\Omega}\right) _{\hat{\zhzeta_i}\hat{\zhzeta_f}}^-}{\left( \dfrac{d\sigma}{d\Omega}\right) _{\hat{\zhzeta_i}\hat{\zhzeta_f}}^++\left( \dfrac{d\sigma}{d\Omega}\right) ^-_{\hat{\zhzeta_i}\hat{\zhzeta_f}}} \label{Sdef1}\,,\\
&&\left( \dfrac{d\sigma}{d\Omega}\right) ^\pm_{\hat{\zhzeta_i}\hat{\zhzeta_f}}=\dfrac{d\sigma}{d\Omega}(\zhp_i,\hat{\zhzeta_i};\zhp_f,\pm\hat{\zhzeta_f}) \label{Sdef2}\,.
\end{eqnarray}
It follows from \Eq{cs} that
\begin{equation}
S_{0\unin}=S_{\unik\unin}=S_{\uniq\unin}=S_{\hat{\zhp_i}\unin}\,, \label{Seq}
\end{equation}
so scattering of electrons polarized along any direction in the plane of scattering (for example, along momentum $\zhp_i$) can also be used to determine $\zhzeta^e$.

Using \Eqsss{zetaD}{zetaS}{Seq} we can write:
\begin{equation}
S_{0\unin}=S_{\unik\unin}=S_{\uniq\unin}=S_{\hat{\zhp_i}\unin}=\dfrac{D(\theta)}{I(\theta)}\label{SD}\,.
\end{equation}

Also, the following expressions can be derived for the remaining functions:
\begin{eqnarray}
S_{\uniq \unik}&=&-S_{\unik \uniq}=\dfrac{F(\theta)}{I(\theta)} \label{SF}\,,\\
S_{\unin \unin}&=&1-\dfrac{H(\theta)+K(\theta)}{I(\theta)+D(\theta)}\,,\\
S_{-\unin-\unin}&=&1-\dfrac{H(\theta)+K(\theta)}{I(\theta)-D(\theta)}\,,\\
S_{\unik \unik}&=&1-\dfrac{G(\theta)+K(\theta)}{I(\theta)}\,,\\
S_{\uniq \uniq}&=&1-\dfrac{G(\theta)+H(\theta)}{I(\theta)}\,.\label{SGH}
\end{eqnarray}

The differential cross section for the scattering of electrons with non-zero projection of polarization on the direction perpendicular to the momentum of the incident electron is dependent not only on the polar angle $\theta$ but also on the azimuthal angle $\varphi$ (the direction of $\unin$ is defined by the plane of scattering and changes with the azimuthal angle $\varphi$ corresponding to the momentum of the scattered electron):
\begin{equation}
\dfrac{d\sigma}{d\Omega}(\theta,\varphi)=\dfrac{1}{4}w I(\theta)\left( 1+\dfrac{D(\theta)}{I(\theta)}(\zhzeta_i\unin)\right)\,. \label{DCSas}
\end{equation}

Now consider the scattering of a polarized electron on an unpolarized ion. Similarly to \Eq{finpol}, the polarization of the scattered electron is given by:
\begin{equation}
\zhzeta^e=\dfrac{\trace(M(1+\zhzeta_i\zhsigma_1)M^+\zhsigma_1)}{\trace(M(1+\zhzeta_i\zhsigma_1)M^+)}\,.
\end{equation}

For polarizations directed along $\unin$ and perpendicular to $\unin$ we get:
\begin{eqnarray}
\zhzeta_i=\pm\unin&\longrightarrow&\zhzeta^e_{\perp}=\pm\unin\left(1-\dfrac{H+K}{I\pm D}\right) =\pm\unin S_{\pm\unin\pm\unin}\,,\\
\zhzeta_i\unin=0&\longrightarrow&\zhzeta^e_{\parallel}=\dfrac{D}{I}\unin + \left( 1-\dfrac{G+K}{I}\right) (\zhzeta_i\unik)\unik + \left( 1-\dfrac{G+H}{I}\right) (\zhzeta_i\uniq)\uniq + \dfrac{F}{I}[\zhzeta_i\times\unin] \nonumber\\
&=&S_{0\unin}\unin+S_{\unik\unik} (\zhzeta_i\unik)\unik+S_{\uniq\uniq}(\zhzeta_i\uniq)\uniq + S_{\uniq\unik}[\zhzeta_i\times\unin], \label{polpos}
\end{eqnarray}
respectively.

The function $F$ determines the change of polarization in the scattering plane, the functions $G$, $H$ and $K$ describe the depolarization of the electron.  

\section{Results and discussion}
The elastic scattering of an electron on an H-like ion can proceed via non-resonant and resonant (with the formation of intermediate autoionizing states) channels. 

We start our discussion by considering the non-resonant channel. If the polarization of electron and ion is of interest it is useful to distinguish between two interactions contributing to this channel: Coulomb interaction of the incident electron with the nucleus (which does not change the polarization of the ion) and its interaction with $1s$ electron (which can result in the spin exchange between two electrons). We note that the Coulomb interaction with the $1s$ electron is partially taken into account nonperturbatively, so that the nucleus is considered to be partially screened by the $1s$ electron.

The amplitude defined by \Eqss{Couldef1}{Couldef2} corresponds to the Coulomb interaction with the partially screened nucleus and describes the Coulomb scattering. The Coulomb scattering of polarized electrons has been thoroughly investigated (see \cite{mott29,mott32,Tolhoek1956,johnson1961}). According to \Eq{zetaD} the polarization of the scattered electron is described by the function $D(\theta)$. If an unpolarized electron is scattered by the Coulomb field of the nucleus, after scattering it has degree of polarization $D(\theta)/I(\theta)$, which then can be observed in the double scattering experiment by looking at the dependency on azimuthal angle of the differential cross section for the second scattering \cite{mott29,mott32}. The effect becomes particularly noticeable for high $Z$. In order to check the accuracy of our calculations for the Coulomb part of the amplitude, we compared our results for $D(\theta)/I(\theta)$ with those obtained in \cite{sherman56}. We found them to be in excellent agreement. A special case of the formula \Eq{cs} for the Coulomb scattering cross section (with $H(\theta) = K(\theta) = 0$) is given in \cite{Tolhoek1956} and \cite{johnson1961}.

\begin{figure}
	\includegraphics[width=0.99\linewidth]{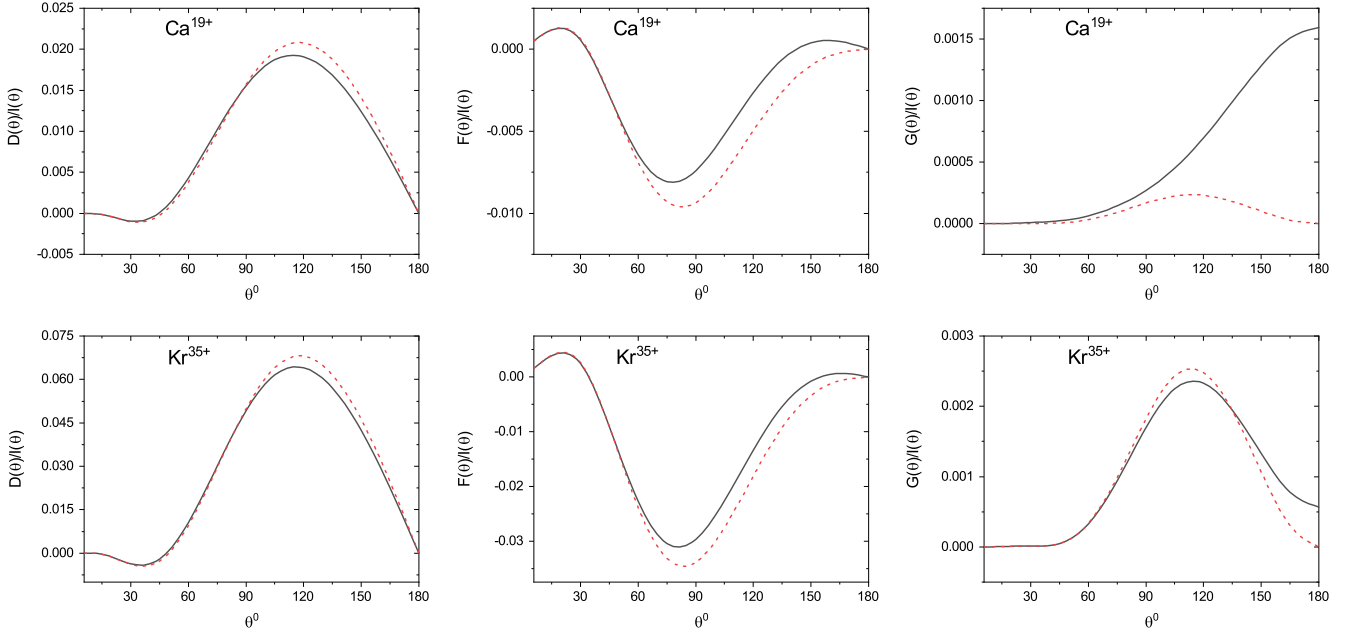}
	\caption{Functions $D(\theta)/I(\theta)$ (first column), $F(\theta)/I(\theta)$ (second column) and $G(\theta)/I(\theta)$(third column) for non-resonant scattering on $\txt{Ca}^{19+}$ and $\txt{Kr}^{35+}$ (black solid line) and Coulomb scattering on the screened potential of the nucleus (red dashed line). The kinetic energy of incident electron is 2.77 keV for $\txt{Ca}^{19+}$ and 9.22 keV for $\txt{Kr}^{35+}$}
	\label{Coulomb}
\end{figure}

Let us first consider the non-resonant energy region of the incident electron. In \Fig{Coulomb} $D/I$, $F/I$ and $G/I$ as functions of the polar angle $\theta$ are presented for the electron scattering on $\txt{Ca}^{19+}$ and $\txt{Kr}^{35+}$. The incident electron kinetic energies (2.77 keV for $\txt{Ca}^{19+}$ and 9.22 keV for $\txt{Kr}^{35+}$) are chosen so that the resonances are far enough to make no contribution. Black solid line corresponds to the scattering on H-like ion and red dashed line represents the results of an approximation in which the role of the $1s$ electron is reduced to screening the nucleus and the exchange interaction between electrons is not taken into account. While for $D/I$ and $F/I$ the contribution due to the exchange interaction with $1s$ electron is rather small and the shape of the curve is mainly determined by the Coulomb interaction, for $G/I$ the situation is different. For small $Z$ the main contribution to the function $G$ comes from the exchange of projections of spin between the incident electron and the $1s$ electron. Thus, removing the spinor structure of the bound electron from consideration (dashed red line) leads to a significant drop in the function $G$. At small $Z$ the main contribution is from the interaction of the incident electron with $1s$ electron and $G/I$ reaches maximum at $180^0$. For higher $Z$ the Coulomb scattering contribution becomes more important. The shape of the curve changes and the maximum moves to around $120^0$.

\begin{figure}
	\includegraphics[width=0.7\linewidth]{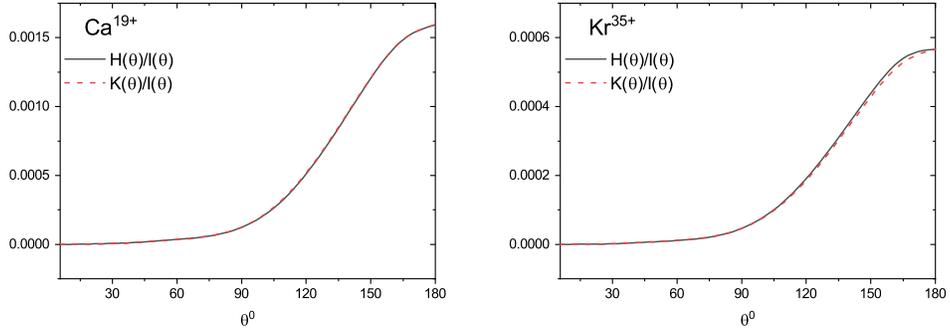}
	\caption{Functions $H(\theta)/I(\theta)$ (black solid line) and $K(\theta)/I(\theta)$ (red dashed line) for non-resonant scattering on $\txt{Ca}^{19+}$ and $\txt{Kr}^{35+}$. The kinetic energy of incident electron is 2.77 keV for $\txt{Ca}^{19+}$ and 9.22 keV for $\txt{Kr}^{35+}$}
	\label{CoulombGHK}
\end{figure}

\Fig{CoulombGHK} presents functions $H/I$ and $K/I$ for the same energies as in \Fig{Coulomb}. For the Coulomb scattering $H(\theta)=K(\theta)=0$, which means that the electron polarization perpendicular to the plane of scattering is not changed by the scattering. The situation is different for the electron scattering on ions with one or more bound electrons. Since the only contribution to $H/I$ and $K/I$ is due to the interaction with the bound electron, these functions decrease with $Z$.

For the non-resonant scattering, the difference between functions $H(\theta)$ and $K(\theta)$ is minimal (especially for small $Z$). Provided that $H(\theta)=K(\theta)$, we can choose $b_{12}=c_{12}$ (see \Eqsss{Mexpansion}{fh}{f2}). Then matrix $M$ defined by \Eq{Mexpansion} is symmetric in respect to the rotation around $\unin$ and the directions $\unik$ and $\uniq$ are no longer special. This approximate symmetry also holds for some of the resonances (see \Fig{Hall} and \Fig{Kall}).

\begin{figure}
	\centering
	\includegraphics[width=0.99\linewidth]{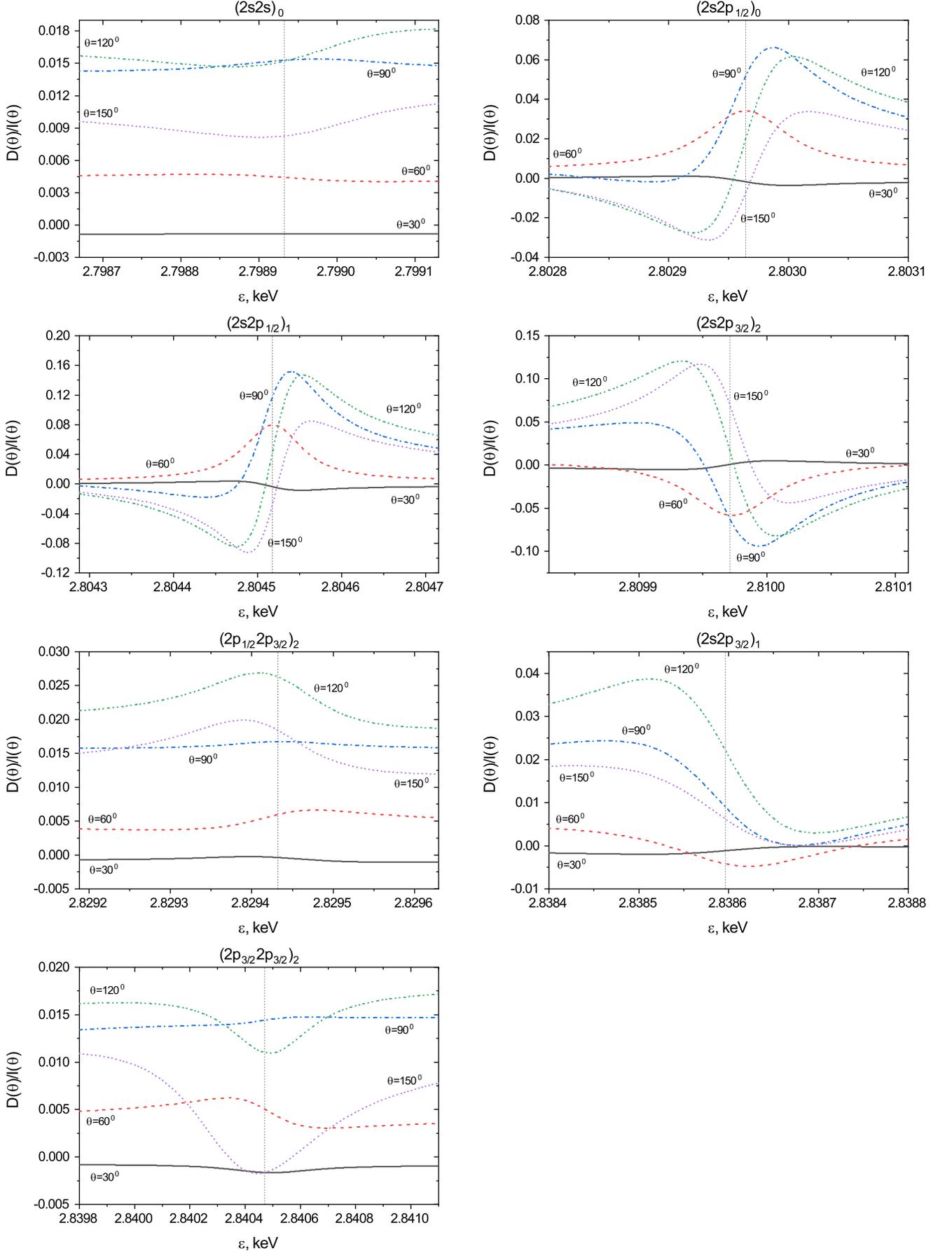}
	\caption{$D(\theta)/I(\theta)$ for electron scattering on $\txt{Ca}^{19+}$ near the resonances, for five different angles $\theta$ of the scattered electron. Vertical grey dashed line denotes the resonance energy.}
	\label{Dall}
\end{figure}

\begin{figure}
	\centering
	\includegraphics[width=0.99\linewidth]{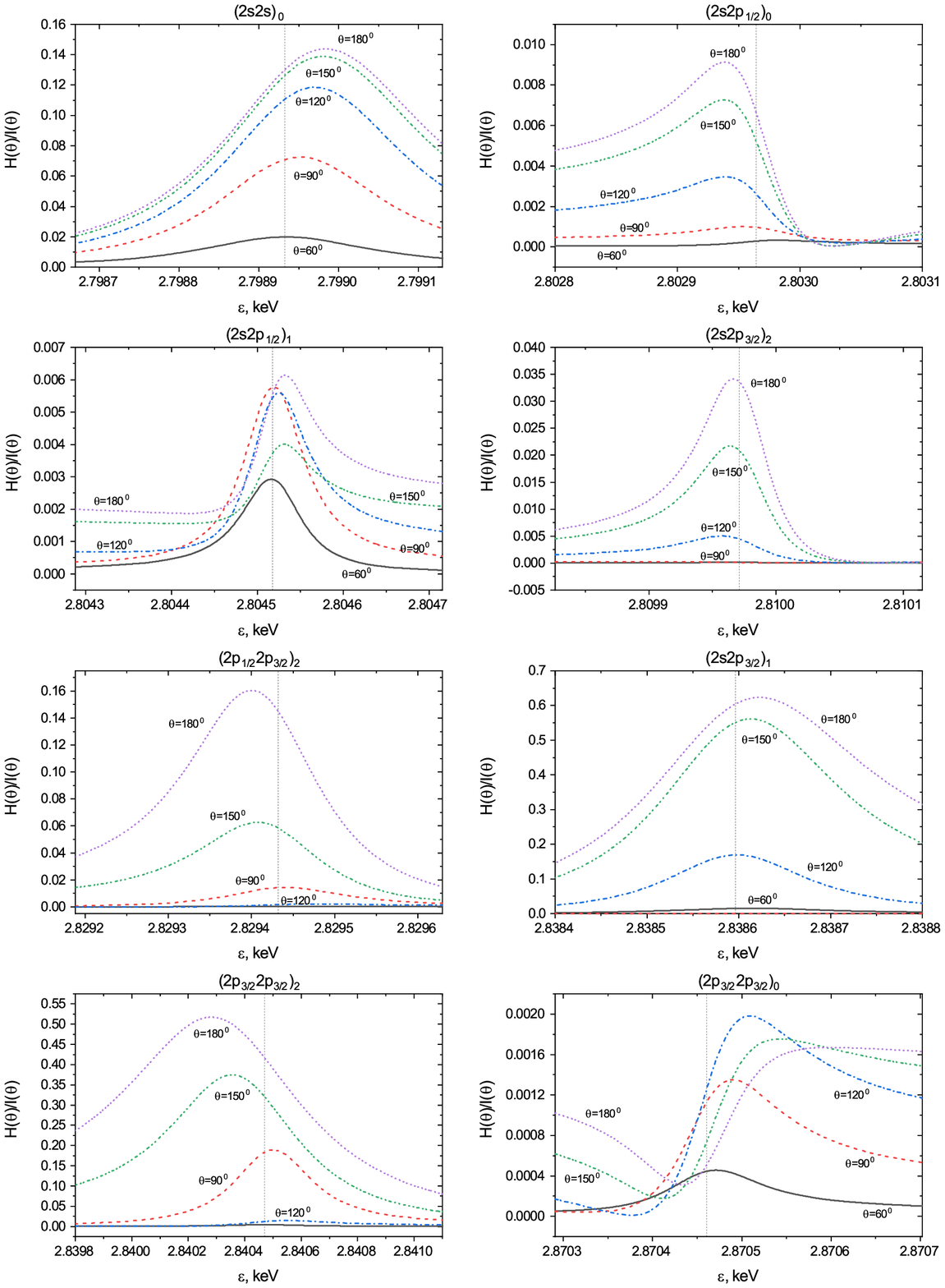}
	\caption{The same as in \Fig{Dall} but for $H(\theta)/I(\theta)$}
	\label{Hall}
\end{figure}

\begin{figure}
	\centering
	\includegraphics[width=0.99\linewidth]{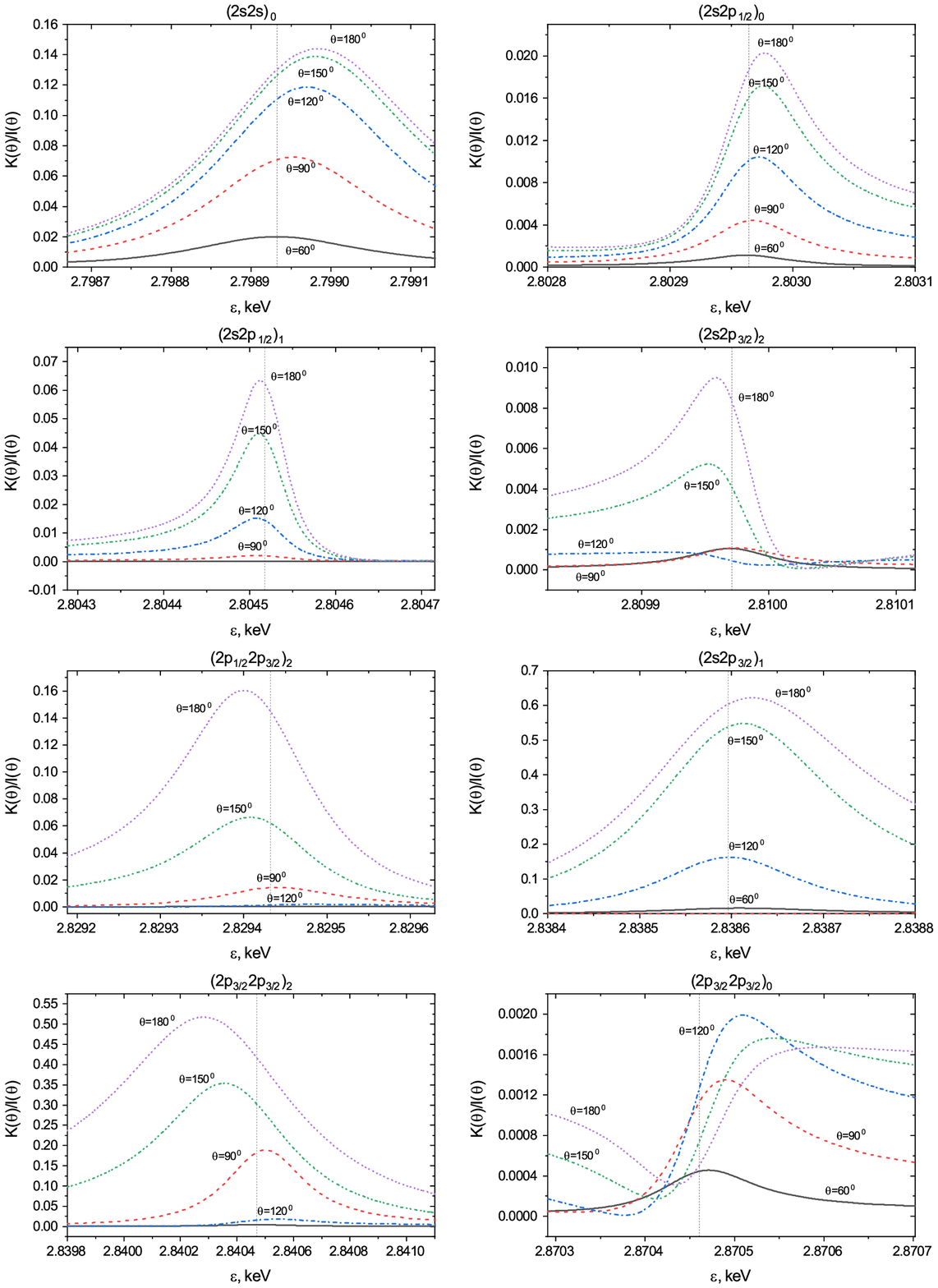}
	\caption{The same as in \Fig{Dall} but for $K(\theta)/I(\theta)$}
	\label{Kall}
\end{figure}

\begin{figure}
	\centering
	\includegraphics[width=0.8\linewidth]{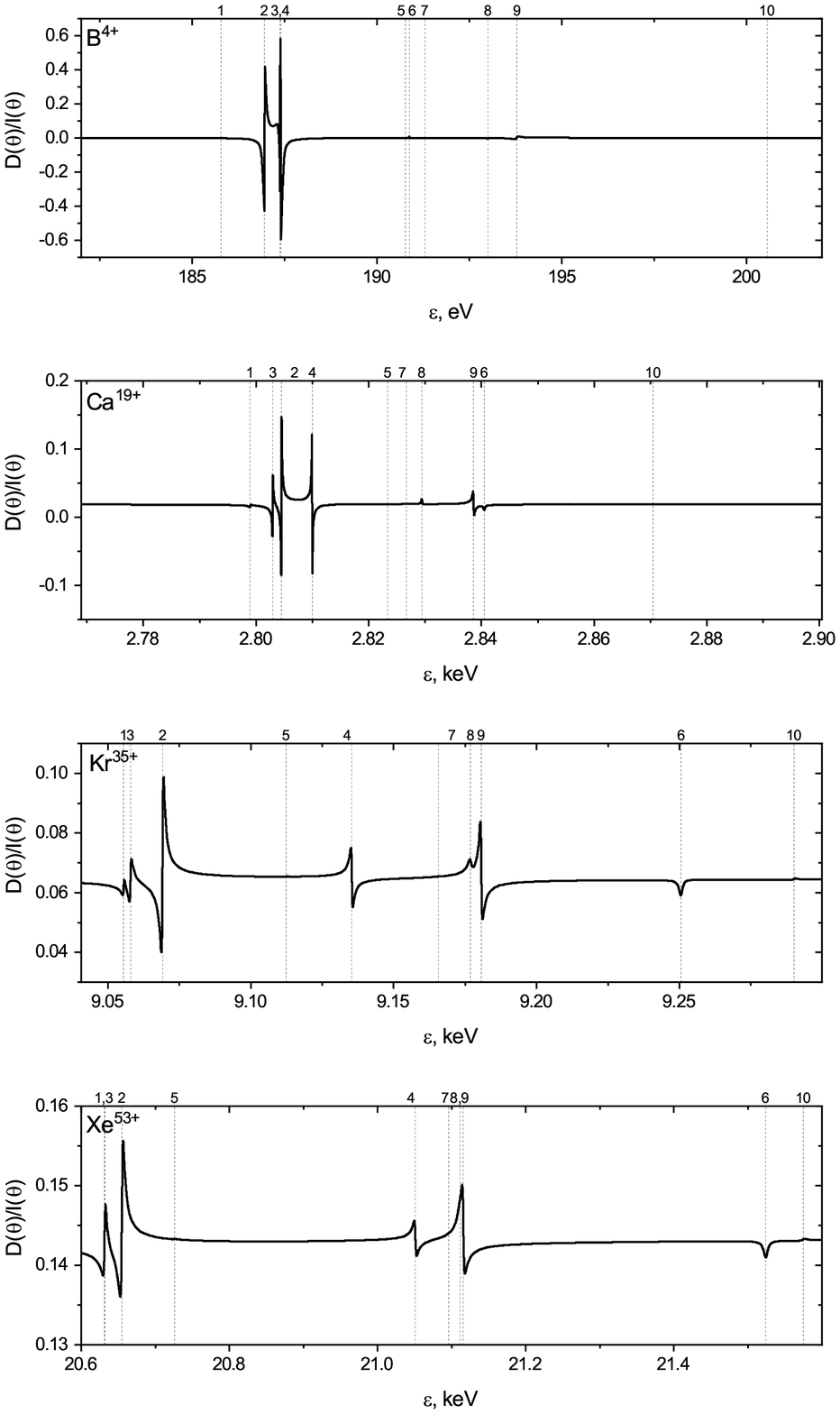}
	\caption{$D/I$ for scattering of electrons on $\txt{B}^{4+}$, $\txt{Ca}^{19+}$, $\txt{Kr}^{35+}$ and $\txt{Xe}^{53+}$ to the angle $\theta=120^{0}$. The vertical dashed lines mark the positions of resonances. The corresponding autoionizing states are presented in Table \ref{resnum}.}
	\label{D120}
\end{figure}

Function $D(\theta)/I(\theta)$ also known as Sherman asymmetry function \cite{sherman56} determines the polarization initially unpolarized electron acquires in the process of scattering (\Eq{finpol}) as well as asymmetry in respect to the azimuthal angle $\varphi$ if the scattered electron has nonzero polarization perpendicular to its momentum (\Eq{DCSas}). Our calculations of $D(\theta)/I(\theta)$ for the scattering on $\txt{Ca}^{19+}$ as a function of the incident electron kinetic energy in the vicinity of the resonances are presented in \Fig{Dall} for five different angles. The resonance shape strongly depends on the scattered electron angle. Depending on the resonance, $D(\theta)/I(\theta)$ reaches maximum at $\theta=90^0-150^0$. Therefore, we chose $\theta=120^0$ to study the behaviour of $D(\theta)/I(\theta)$ for the scattering on different ions. Parameter $D/I$ at $\theta=120^0$ for the electron scattering on $\txt{B}^{4+}$, $\txt{Ca}^{19+}$, $\txt{Kr}^{35+}$ and $\txt{Xe}^{53+}$ is presented in \Fig{D120} as a function of the incident electron kinetic energy. With the growth of $Z$, the role of the background becomes more significant.

\begin{figure}
	\centering
	\includegraphics[width=0.99\linewidth]{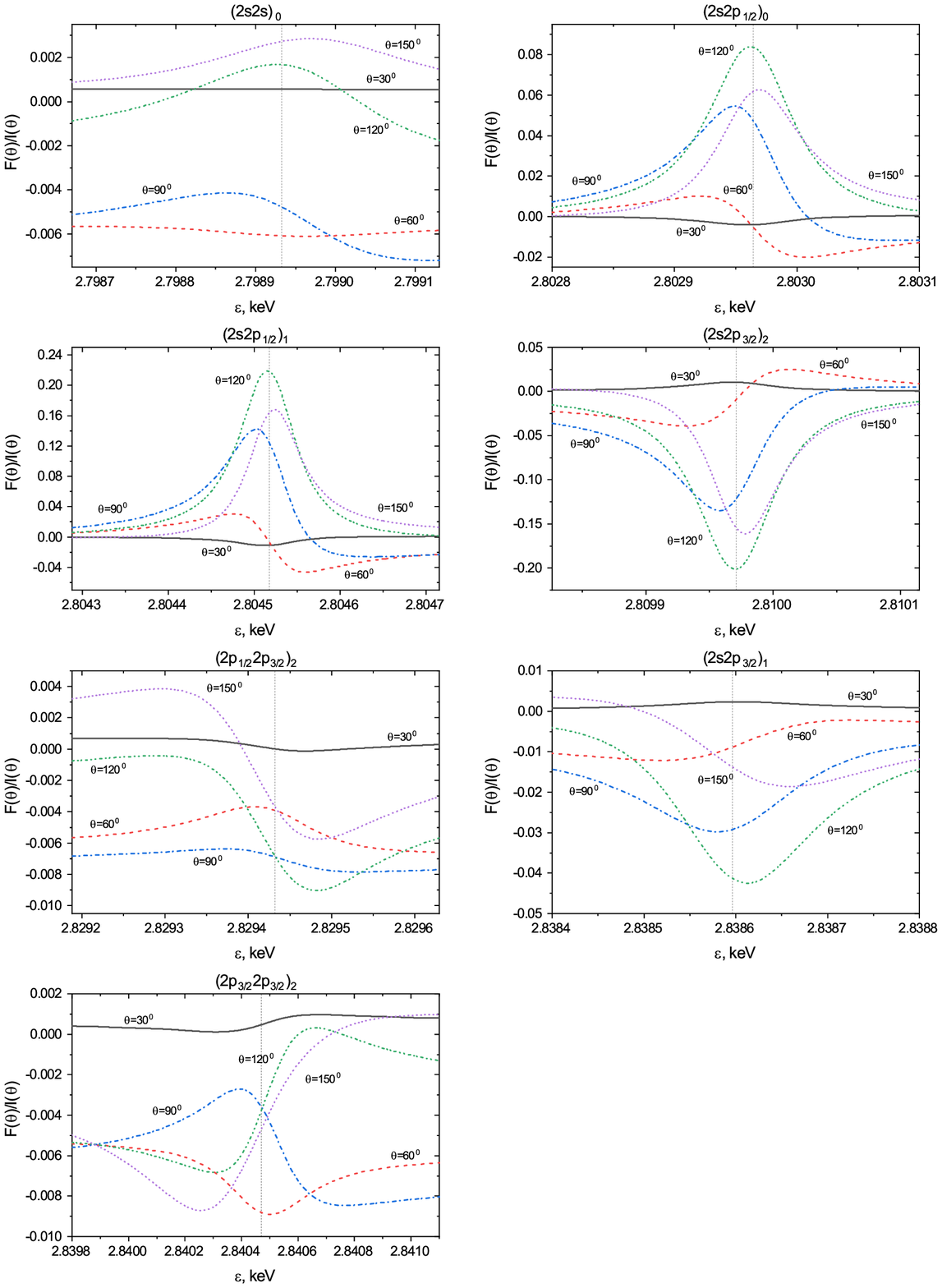}
	\caption{The same as in \Fig{Dall} but for $F(\theta)/I(\theta)$}
	\label{Fall}
\end{figure}

\begin{figure}
	\centering
	\includegraphics[width=0.8\linewidth]{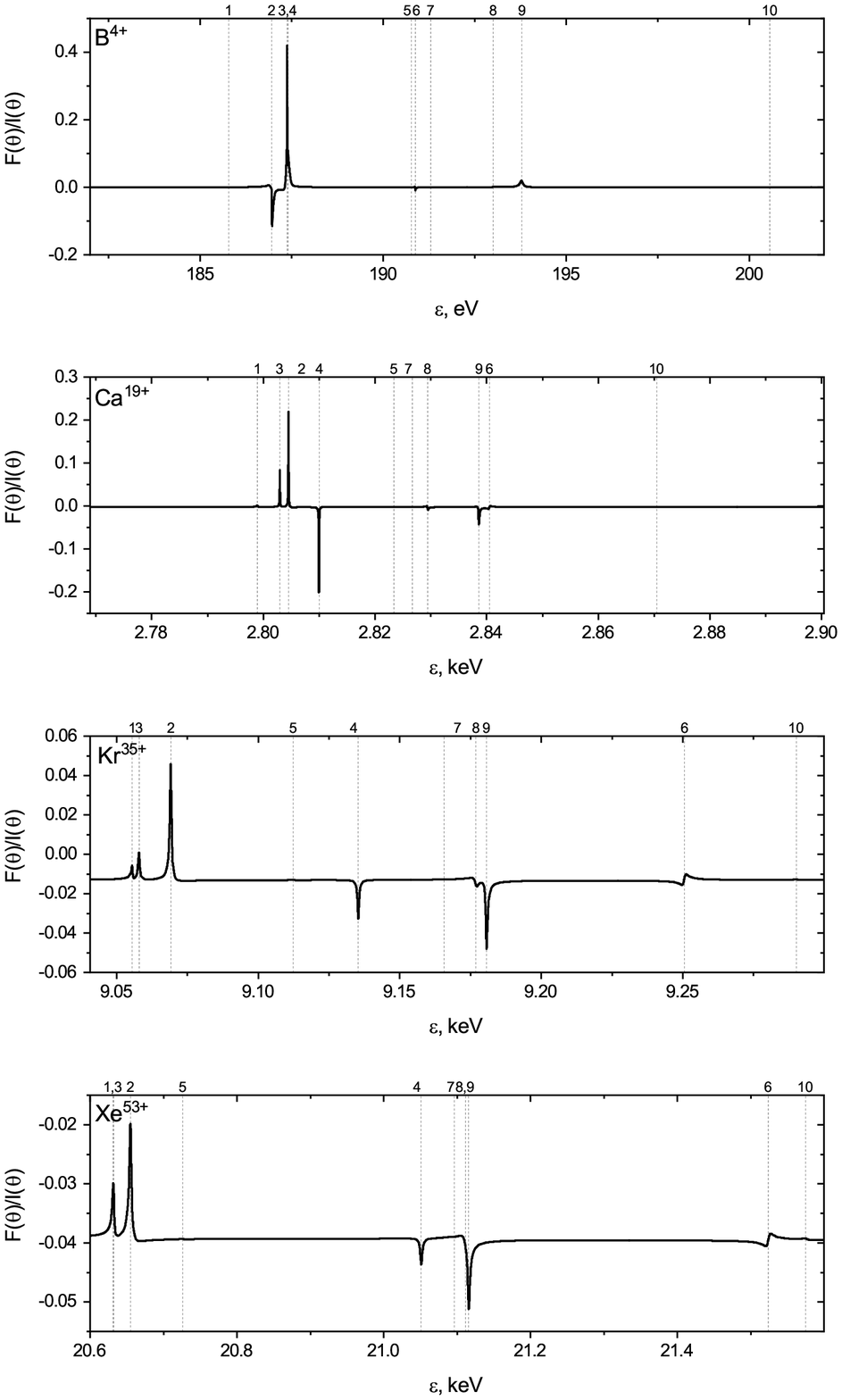}
	\caption{The same as in \Fig{D120} but for $F/I$}
	\label{F120}
\end{figure}

In Figs. \ref{Fall} and \ref{F120} results for $F(\theta)/I(\theta)$ are presented in the same manner. The physical meaning of the function $F/I$ is as follows. If the incident electron is longitudinally polarized, $F(\theta)/I(\theta)$ determines the scattered electron polarization component lying in the plane of scattering and perpendicular to the incident electron momentum.

For the description of depolarization of electron in the process of scattering, it is convenient to alternate between functions $G(\theta)/I(\theta)$, $H(\theta)/I(\theta)$, $K(\theta)/I(\theta)$ (\Eqsamb{fg}{f2}) and parameters $S_{\unik\unik}$, $S_{\uniq\uniq}$ and $S_{\unin\unin}$. Parameters $S_{\hat{\zhzeta}\hat{\zhzeta}}$ describe the scattering of an electron with $+1/2$ spin projection on the direction $\hat{\zhzeta}$ and represent the difference between the number of electrons that have the spin projection $+1/2$ and $-1/2$ on the direction $\hat{\zhzeta}$ after scattering normalized by the total number of scattered electrons (\Eqss{Sdef1}{Sdef2}). Parameters $S_{\unik\unik}$, $S_{\uniq\uniq}$ and $S_{\unin\unin}$ directly correspond to the change of projection of spin on certain direction for three special directions while functions $G(\theta)/I(\theta)$, $H(\theta)/I(\theta)$, $K(\theta)/I(\theta)$ have clear connection to coefficients $b_{12}$ and $c_{12}$ from \Eq{Mexpansion}. The connection between $G(\theta)/I(\theta)$, $H(\theta)/I(\theta)$ and $K(\theta)/I(\theta)$ and depolarization tensor introduced in \cite{Burke1974} is presented in Appendix \ref{diffnot}.

In Figs. \ref{Gall}, \ref{Hall} and \ref{Kall} functions $G(\theta)/I(\theta)$, $H(\theta)/I(\theta)$ and $K(\theta)/I(\theta)$ are presented for energies close to the resonances and for five different scattered electron angles. For most resonances the maxima of these functions are reached at $\theta=180^0$. 

\begin{figure}
	\centering
	\includegraphics[width=0.8\linewidth]{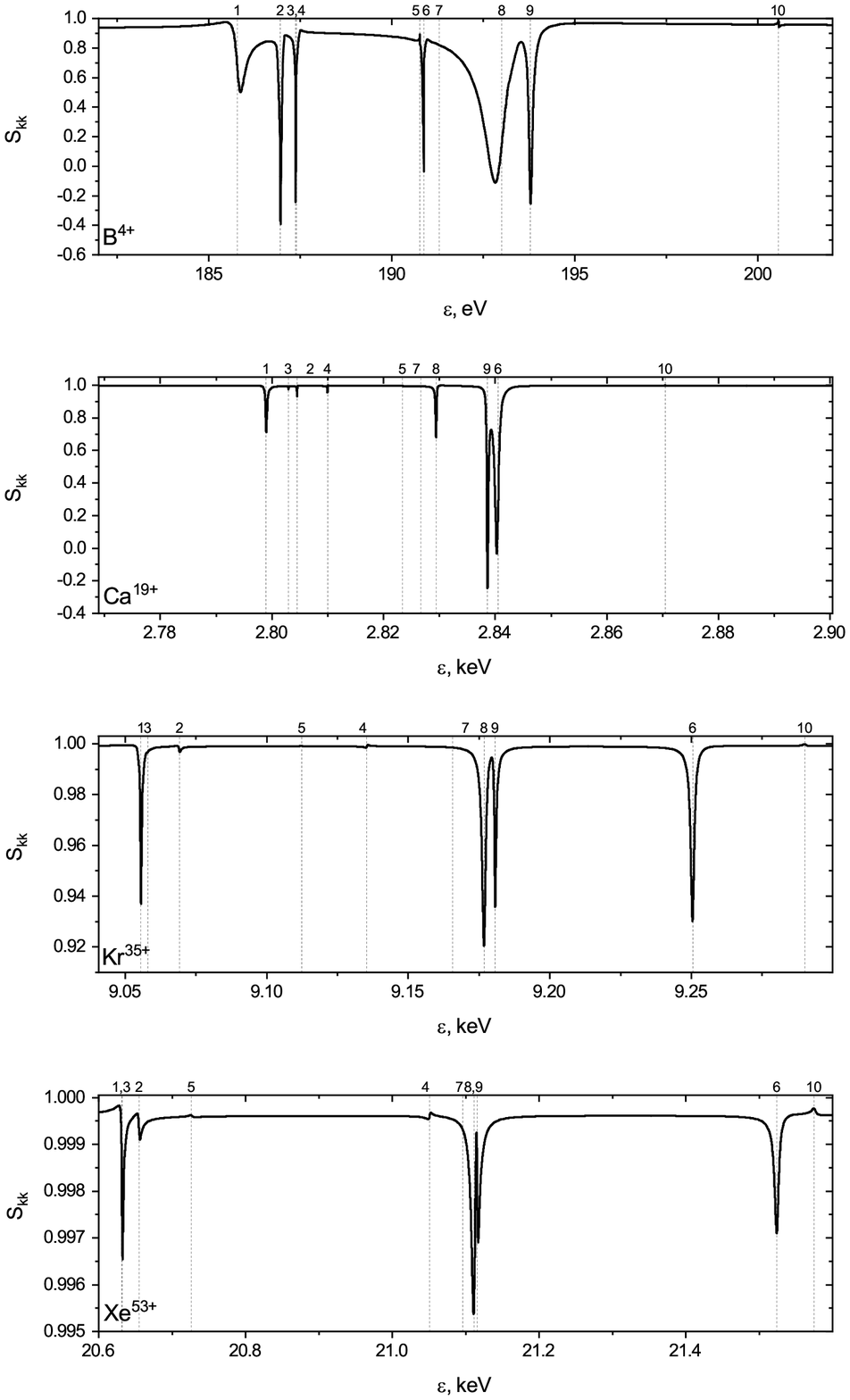}
	\caption{$S_{\unik\unik}$ for scattering of electrons on $\txt{B}^{4+}$, $\txt{Ca}^{19+}$, $\txt{Kr}^{35+}$ and $\txt{Xe}^{53+}$ to the angle $\theta=180^{0}$. The vertical dashed lines mark the positions of resonances. The corresponding autoionizing states are presented in Table \ref{resnum}.}
	\label{Skk180}
\end{figure}

\begin{figure}
	\centering
	\includegraphics[width=0.8\linewidth]{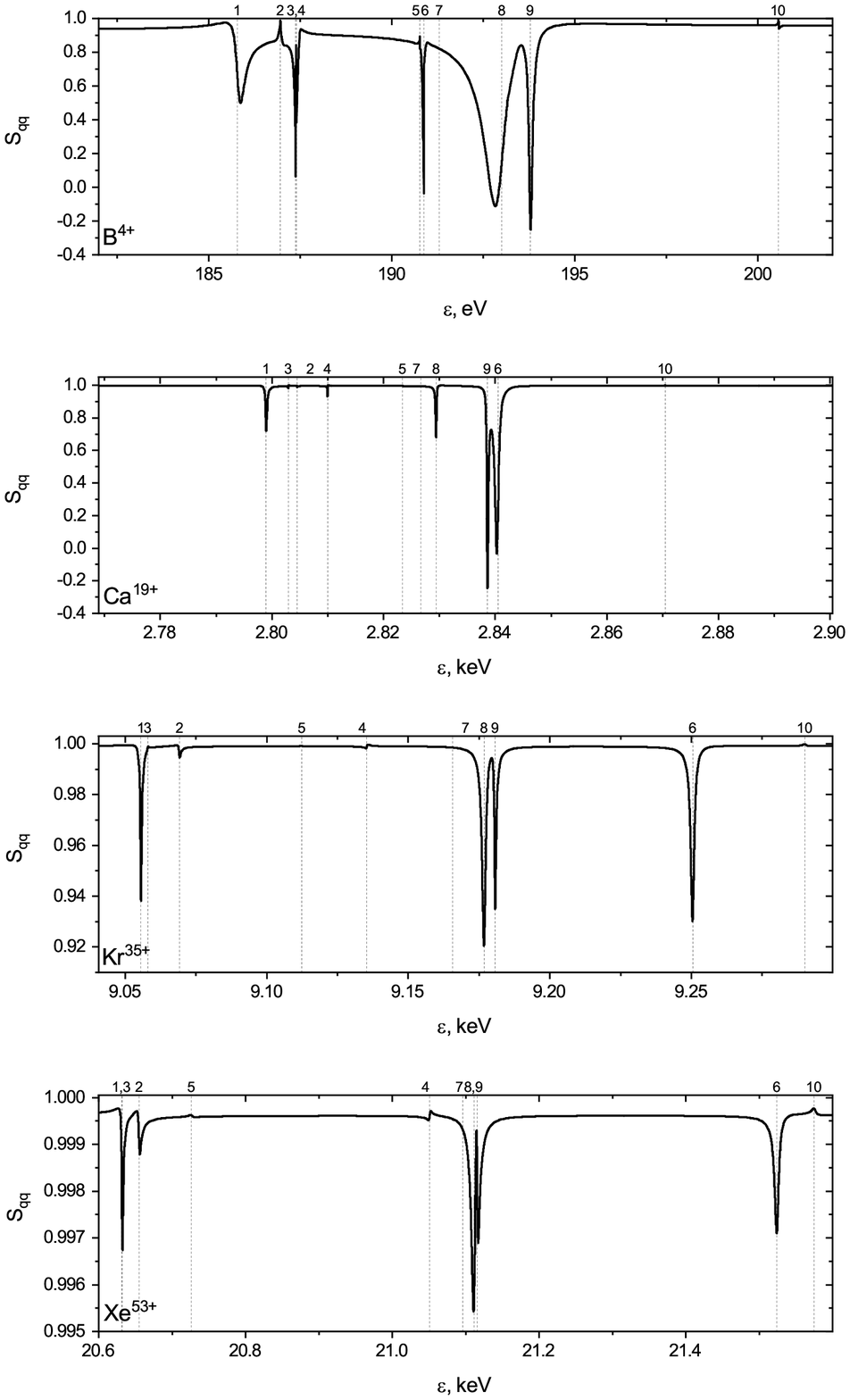}
	\caption{The same as in \Fig{D120} but for $S_{\uniq\uniq}$}
	\label{Sqq180}
\end{figure}

In Figs. \ref{Skk180} and \ref{Sqq180} $S_{\unik\unik}(180^0)=S_{\unin\unin}(180^0)$ and $S_{\uniq\uniq}(180^0)$ are presented for scattering on $\txt{B}^{4+}$, $\txt{Ca}^{19+}$, $\txt{Kr}^{35+}$ and $\txt{Xe}^{53+}$. These figures show that the probability of a change in the spin projection is highly dependent on the specific resonance.

\begin{figure}
	\centering
	\includegraphics[width=0.4\linewidth]{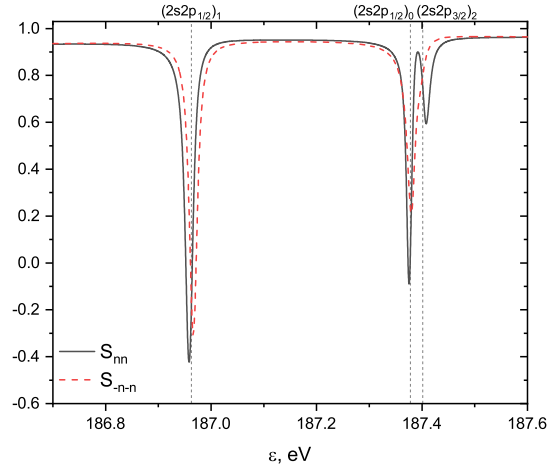}
	\caption{$S_{\unin\unin}$ (black solid line) and $S_{-\unin-\unin}$ (dashed red line) for scattering on $\txt{B}^{4+}$ to the angle $\theta=120^0$ in the vicinity of $(2s2p_{1/2})_1$, $(2s2p_{1/2})_0$ and $(2s2p_{3/2})_2$ resonances}
	\label{Snnpm120}
\end{figure}

\Fig{Snnpm120} shows the depolarization asymmetry for the scattering of electrons on $\txt{B}^{4+}$ with opposite polarizations along $\unin$. The asymmetry manifests in the difference between $S_{\unin\unin}$ and $S_{-\unin-\unin}$. The incident electron energy and the scattering angle were chosen to maximise the asymmetry function $D/I$.  

\begin{figure}
	\centering
		\includegraphics[height=0.37\linewidth]{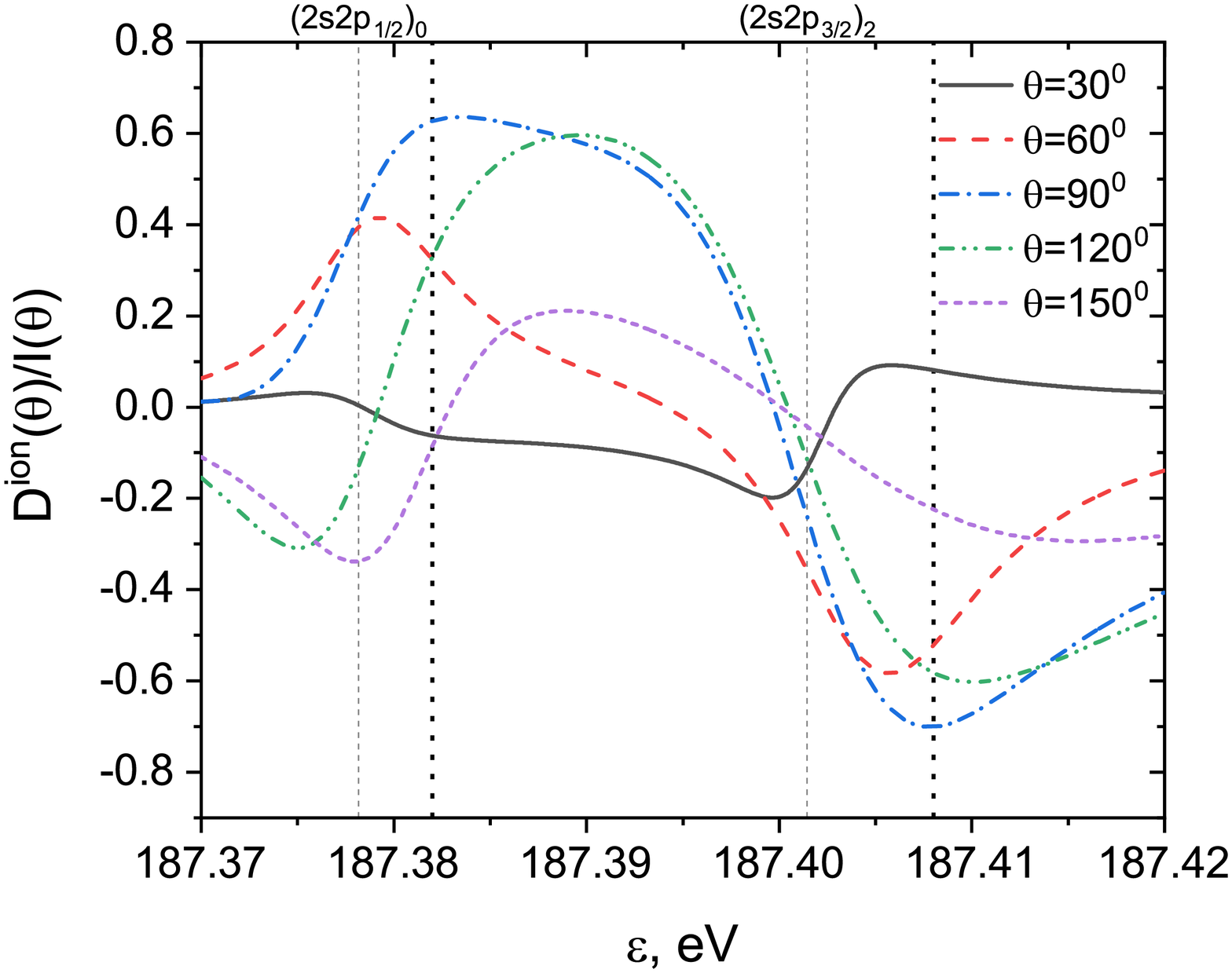}
		\includegraphics[height=0.37\linewidth]{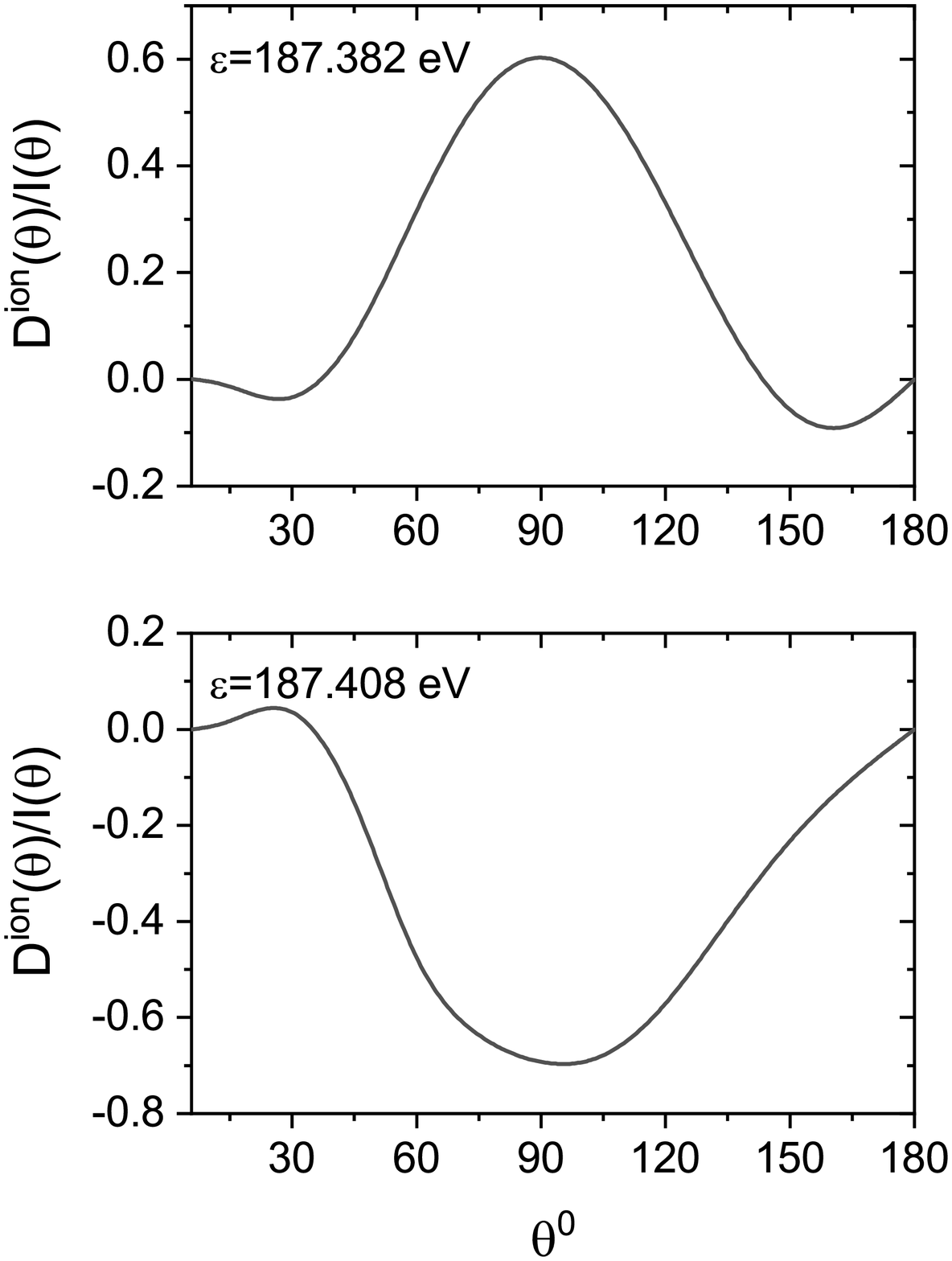}
	\caption{$D^{\txt{ion}}(\theta)/I(\theta)$ for scattering on $\txt{B}^{4+}$ as function of incident electron energy $\varepsilon$ in the vicinity of $(2s2p_{1/2})_0$ and $(2s2p_{3/2})_2$ resonances (graph on the left) and as function of $\theta$ at energies marked by black dotted lines on the left graph (graphs on the right)}
	\label{ionpolar}
\end{figure}

In the collisions of an unpolarized electron with an unpolarized ion the ion also acquires polarization perpendicular to the plane of scattering equal to $D^{\txt{ion}}(\theta)/I(\theta)$. In \Fig{ionpolar} function $D^{\txt{ion}}(\theta)/I(\theta)$ is presented for the electron scattering on $\txt{B}^{4+}$ as a function of the incident electron kinetic energy (on the left) and as a function of the scattered electron angle (on the right). The polarization of an initially unpolarized ion after collision can reach $70\%$.

Function $D^{\txt{ion}}(\theta)/I(\theta)$ also determines the cross section asymmetry in respect to the azimuthal angle $\varphi$. If an unpolarized electron is scattered on a polarized ion, the differential cross section is dependent on both polar and azimuthal angles. If the polarization of the ion $\zh{\eta}$ is given by the degree of polarization $P$, the polar angle $\chi$ and the azimuthal angle $\omega$, the cross section can be written as:
\begin{equation}
\dfrac{d\sigma}{d\Omega}(\theta,\varphi)\sim I(\theta)+D^{\txt{ion}}(\theta)P\sin\chi\sin(\omega-\varphi)\,.\label{polioncs}
\end{equation}
This formula is completely analogous to \Eq{DCSas} with $\zhzeta$ and $D(\theta)$ switched to $\zh{\eta}$ and $D^{\txt{ion}}(\theta)$.
 
By looking at the asymmetry of the cross section in respect to $\varphi$ one can determine the ion polarization component transverse to the incident electron momentum. If $\theta$ is fixed then the maximum and the minimum of the differential cross sections are reached at $\varphi=\omega-\pi/2$ and $\varphi=\omega+\pi/2$ respectively and the difference between the maximum and the minimum is equal to $2D^{\txt{ion}}(\theta)P_\perp$, where $P_\perp$ is the ion polarization component transverse to the incident electron momentum. Similar principle is used in Mott polarimetry \cite{Gay1992}.
\section{Conclusion}
 In this paper we investigated the polarization correlations for the elastic electron scattering on H-like ions. For this purpose, we utilized the scattering matrix $M$ (\Eq{Mexpansion}) which completely determines the cross section in terms of the polarizations of ions and incident electrons. In order to calculate $M$ we employed the relativistic QED theory developed in \cite{res2020}. We performed our research for a wide range of ions (from B$^{4+}$ to Xe$^{53+}$). Special attention was paid to the regions of the impact energies at which intermediate LL autoionizing states can be formed.
 
 Our main focus was on the scattering of electrons on unpolarized ions. The electron polarization change in this process can be completely described by the five parameters $D/I$, $F/I$, $G/I$, $H/I$ and $K/I$. These five parameters were thoroughly investigated for both the non-resonant and resonant energy region as functions of the incident electron energy and the scattered electron angle. 
  
 For the resonant electron scattering on ions with $Z$ up to 50 both the relativistic spin-orbit interaction and the exchange interaction are significant and the polarization properties are determined by the combination of these two effects.  For higher $Z$ the spin-orbit interaction becomes dominant. The importance of the exchange interaction is clearly manifested in the behaviour of the parameters $H$ and $K$. In the absence of the exchange interaction ($H=K=0$) the electron with the polarization perpendicular to the plane of scattering does not change polarization in the scattering. The situation is drastically different for the resonant electron scattering on H-like ions. In some cases the electron polarized perpendicular to the plane of scattering not only can change its polarization to the opposite after the scattering but also is more likely to do so.
 
 In the scattering of an unpolarized electron on an unpolarized ion both the electron and the ion become partially polarized. At some resonances the initially unpolarized electron can gain polarization up to 60\% after scattering on B$^{4+}$ and up to 15\% after scattering on Ca$^{19+}$.  
 
 Since the polarization phenomena caused by the presence of resonances are more pronounced for lighter ions, we considered the resonant electron scattering on B$^{4+}$ in order to estimate the polarization that an initially unpolarized ion can obtain in the process of scattering. For this purpose we calculated the parameter $D^{\txt{ion}}(\theta)/I(\theta)$. This parameter determines the polarization that initially unpolarized ion acquires in the process of scattering as well as the asymmetry of the differential cross section for the scattering on polarized ions. It is the analogue of the Sherman asymmetry function for ions.  We found that for B$^{4+}$ the asymmetry is significant and for certain impact energies and scattering angles can reach 70\%.  As in the case of electrons, this asymmetry can be used for producing or measuring ion polarization.
\begin{acknowledgments}
	The work of D.M.V. was supported by the Dmitrij Mendeleev-Programm 2020 (FPI 57516244).
	The work of D.M.V. and O.Y.A. on the calculation of the differential cross section was supported solely by the Russian Science Foundation under Grant No. 17-12-01035.
	The reported study was funded by RFBR and NSFC according to the research project No. 20-52-53028.
	This work is supported by the National Key Research and Development Program of China under Grant No. 2017YFA0402300, the National Natural Science Foundation of China under Grant No. 11774356, the Chinese Academy of Sciences (CAS) President's International Fellowship Initiative (PIFI) under Grant No. 2018VMB0016 and Chinese Postdoctoral Science Foundation No. 2020M673538.
\end{acknowledgments}
\appendix
\section{}
\label{ffs}
For the sake of convenience, we choose the indices of matrix $M$ to be in order $(++,+-,-+,--)$, where the first index refers to the polarization $\mu$ and the second index refers to the spin projection to $z$-axis $m$ of the bound electron. Also we fix the initial polarization $\mu_i$ as $+1/2$. Then the corresponding eight matrix elements can be written as:
\begin{eqnarray}
M_{11}=U_{m_i=+1/2,\mu_i=+1/2;m_f=+1/2,\mu_f=+1/2}\,,\\
M_{21}=U_{m_i=+1/2,\mu_i=+1/2;m_f=-1/2,\mu_f=+1/2}\,,\\
M_{31}=U_{m_i=+1/2,\mu_i=+1/2;m_f=+1/2,\mu_f=-1/2}\,,\\
M_{41}=U_{m_i=+1/2,\mu_i=+1/2;m_f=-1/2,\mu_f=-1/2}\,,\\
M_{12}=U_{m_i=-1/2,\mu_i=+1/2;m_f=+1/2,\mu_f=+1/2}\,,\\
M_{22}=U_{m_i=-1/2,\mu_i=+1/2;m_f=-1/2,\mu_f=+1/2}\,,\\
M_{32}=U_{m_i=-1/2,\mu_i=+1/2;m_f=+1/2,\mu_f=-1/2}\,,\\
M_{42}=U_{m_i=-1/2,\mu_i=+1/2;m_f=-1/2,\mu_f=-1/2}\,.
\end{eqnarray}

If these matrix elements are known, the remaining eight can be easily found since:
\begin{equation}
U_{m_i,\mu_i;m_f,\mu_f}=U_{-m_i,-\mu_i;-m_f,-\mu_f}e^{2i(m_i+\mu_i-m_f-\mu_f)\varphi}(-1)^{m_i+\mu_i-m_f-\mu_f}\,.
\end{equation}

Then for the coefficients in the expansion of $M$ (\Eq{Mexpansion}) we can write:
\begin{eqnarray}
a_0&=&\dfrac{M_{11}+M_{22}}{2}\,,\\
a_1&=&\dfrac{M_{31}+M_{42}}{2i}e^{-i\varphi}\,,\\
a_2&=&\dfrac{M_{21}e^{-i\varphi}-M_{12}e^{i\varphi}}{2i}\,,\\
a_{12}&=&\dfrac{M_{32}-M_{41}e^{-2i\varphi}}{2}\,,\\
b_{12}&=&(1+\sec\theta)\dfrac{M_{11}-M_{22}}{4}+(1-\sec\theta)\dfrac{M_{41}e^{-2i\varphi}+M_{32}}{4}\\
&=&\dfrac{M_{11}-M_{22}}{2}+\tan\dfrac{\theta}{2}\dfrac{M_{21}e^{-i\varphi}+M_{12}e^{i\varphi}}{2}\,,\\
c_{12}&=&(1-\sec\theta)\dfrac{M_{11}-M_{22}}{4}+(1+\sec\theta)\dfrac{M_{41}e^{-2i\varphi}+M_{32}}{4}\\
&=&\dfrac{M_{11}-M_{22}}{2}-\cot\dfrac{\theta}{2}\dfrac{M_{21}e^{-i\varphi}+M_{12}e^{i\varphi}}{2}\,.
\end{eqnarray}

Also, since $M$ can be described with 6 complex numbers, out of 8 matrix elements defined above only six are independent:
\begin{eqnarray}
&&M_{42}-M_{31}+M_{21}+M_{12}e^{2i\varphi}=0\\
&&\sin\theta(M_{11}-M_{22}-M_{32}-M_{41}e^{-2i\varphi})-2\cos\theta(M_{21}e^{-i\varphi}+M_{12}e^{i\varphi})=0
\end{eqnarray}

The additional symmetry at $\theta=\pi$ results in additional limitations on functions $D$, $F$, $G$ and $H$ at this angle. Consider backward scattering ($\theta=\pi$, $\hat{\zhp_i}=\uniq$). If the incident electron is polarized along the direction of its momentum, the process is symmetric in respect to the $z$-axis, which means that cross section does not depend on the electron polarization perpendicular to the $z$-axis. Therefore, $S_{\hat{\zhp_i}\unin}=S_{\hat{\zhp_i}\unik}=0$ and $S_{\unin \unin}=S_{\unik \unik}$. This, along with the Eqs.\ (\ref{Sdef1}-\ref{Sdef2}, \ref{SD}-\ref{SGH}), immediately leads to:
\begin{eqnarray}
D(\pi)=0,\: F(\pi)=0,\: G(\pi)=H(\pi) \,.
\end{eqnarray}

\section{}\label{diffnot}
In previous investigations of the electron scattering a number of parameters were introduced to characterize the polarization properties. In this paper we adapted the set of parameters used to describe the Coulomb scattering in \cite{johnson1961} to the electron scattering on H-like ions. 

The following formulae can be used to switch to the notations introduced by Burke for the electron scattering on H-like ions \cite{Burke1974}:
\begin{eqnarray}
I_0&=&I\,,\\
P&=&\dfrac{D}{I}\,,\\
D_{kq}&=&\dfrac{F}{I}\,,\\
D_{nn}&=&1-\dfrac{H+K}{I}\,,\\
D_{kk}&=&1-\dfrac{G+K}{I}\,,\\
D_{qq}&=&1-\dfrac{G+H}{I}\,.
\end{eqnarray}

In order to describe the polarization properties of the electron scattering on atoms, parameters $S$, $T$ and $U$ are usually introduced \cite{Dapor2018}. While for the resonant electron scattering on H-like ions the use of only three parameters is clearly insufficient, we can still establish the connection between $S$, $T$ and $U$ and the parameters introduced in this paper. If we neglect the spin exchange between the incident and bound electrons, parameters $H(\theta)$ and $K(\theta)$ become zero. In that case,
\begin{eqnarray}
S(\theta)&=&\dfrac{D(\theta)}{I(\theta)}\,,\\
T(\theta)&=&1-\dfrac{G(\theta)}{I(\theta)}\,,\\
U(\theta)&=&-\dfrac{F(\theta)}{I(\theta)}\,,
\end{eqnarray}
and
\begin{equation}
S^2+T^2+U^2=1\,. \label{fpolnorm}
\end{equation}
\begin{figure}
	\centering
	\includegraphics[width=0.99\linewidth]{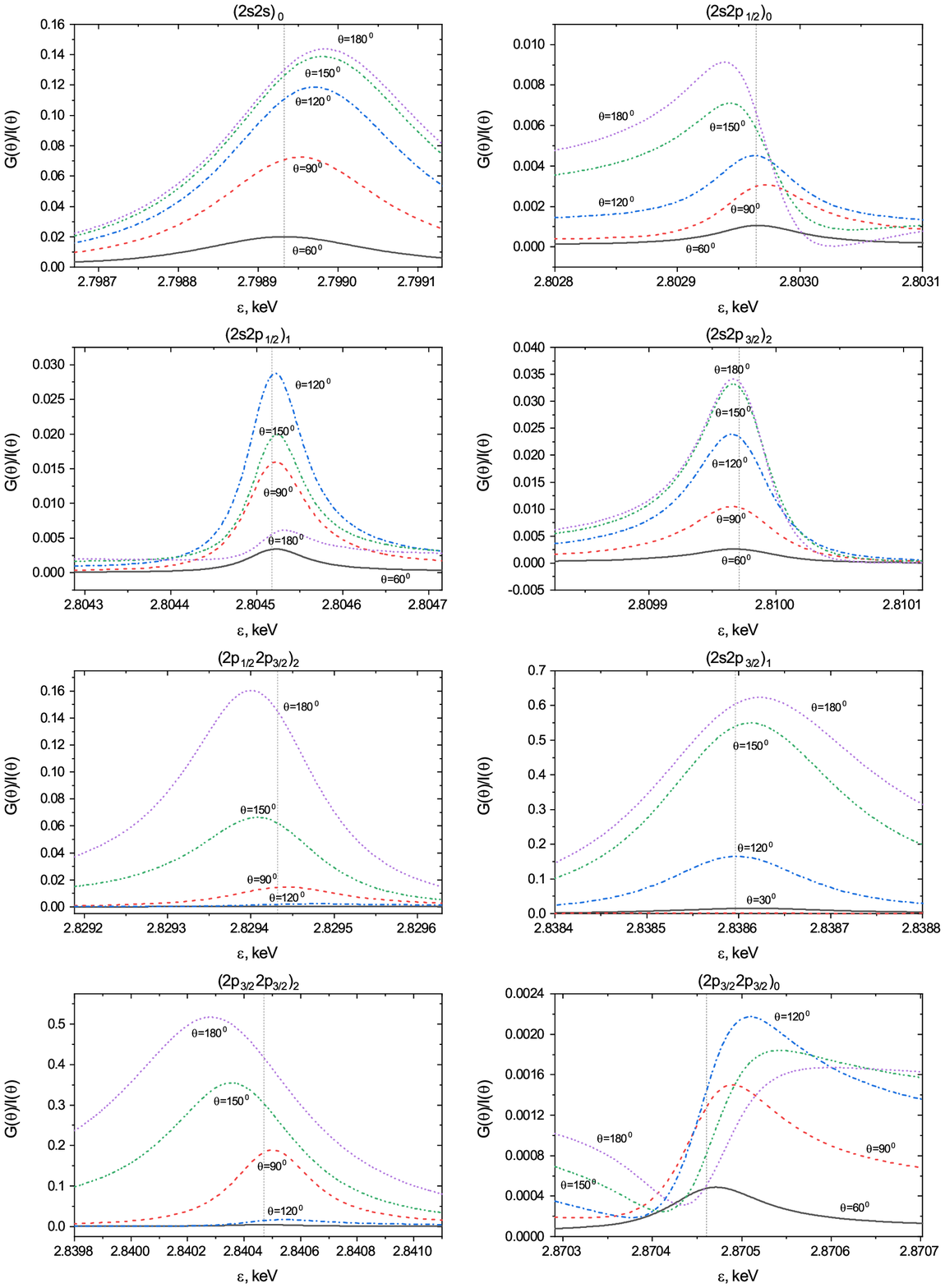}
	\caption{The same as in \Fig{Dall} but for $G(\theta)/I(\theta)$}
	\label{Gall}
\end{figure}
\begin{table}	
	\caption{Order of resonances in Figs. \ref{D120} - \ref{Sqq180}}
	\label{resnum}
	\begin{tabular}{cc}
		\hline
		resonance&intermediate\\
		number&autoionizing state\\
		\hline
		1&$(2s)_0^2$\\
		2&$(2s2p_{1/2})_1$\\
		3&$(2s2p_{1/2})_0$\\
		4&$(2s2p_{3/2})_2$\\
		5&$(2p_{1/2})_0^2$\\
		6&$(2p_{3/2})_2^2$\\
		7&$(2p_{1/2}2p_{3/2})_1$\\
		8&$(2p_{1/2}2p_{3/2})_2$\\
		9&$(2s2p_{3/2})_1$\\
		10&$(2p_{3/2})_0^2$\\
		\hline
	\end{tabular}
\end{table}

\end{document}